\documentclass[]{IEEEtran}

\usepackage[T1]{fontenc}
\usepackage{lmodern}

\usepackage{ucs}

\usepackage[english]{babel}
\usepackage{euscript}
\usepackage{amstext}
\usepackage{mathrsfs,amssymb}
\usepackage{amsmath}  
\usepackage{amsfonts}
\usepackage{makeidx} 
\usepackage{url}
\usepackage{color}
\usepackage{pslatex}
\usepackage{comment}
\usepackage[english]{diffs}
\usepackage{xspace}
\usepackage{listings}
\usepackage{subfigure}
\usepackage[pdftex]{graphicx}
\usepackage{./kermeta}
\usepackage{./emftext}

\usepackage{textpos}

\usepackage{hyperref} 

\usepackage[right]{eurosym}
\usepackage{theorem}

\usepackage{color}
\definecolor{DBlue}{rgb}{0,0,0.4}
\definecolor{DRed}{rgb}{0.4,0,0}
\definecolor{CGreen}{rgb}{0.3,0.7,0.4}
\definecolor{OliveGreen}{rgb}{0.3,0.7,0.4}
\definecolor{LGrey}{rgb}{0.95,0.95,0.95}
\usepackage[]{listings}
\lstdefinelanguage{scala}{
       morekeywords={
                override, try, catch, throw, private, public, protected, import, package, implicit, final, package, trait, type, class, val, def, var, if, this, else, extends, with, while, new, abstract, object, case, match, sealed, for, yield,super},
         sensitive=t, 
   morecomment=[s]{/*}{*/},morecomment=[l]{\//},
   escapeinside={/*\%}{*/},
   rangeprefix= /*< ,rangesuffix= >*/,
   morestring=[d]{"}
 }
\lstdefinelanguage{kermeta}
{
morekeywords=[1]{ package, inherits, class, abstract, attribute, reference, property, readonly, getter, is, setter, alias, enumeration, operation, method },
morekeywords=[4]{ bag, set, seq, oset, Void },
morekeywords=[5]{ and, or, not },
morekeywords=[2]{ do, end, function, rescue, from, until, loop, if, then, else, raise, var, init, raises },
morekeywords=[3]{ true, false, void, result, value, self, super },
morekeywords=[0]{ require, using, extern, pre, post, inv },sensitive=true,
morecomment=[l]{//},
morecomment=[s]{/*}{*/},
morestring=[b]",
basicstyle=\footnotesize
}

\usepackage{enumitem}

\usepackage{tikz}
\usepackage{tikz-uml}
\usepackage{tikz-timing}
\usetikzlibrary{calc}
\usetikzlibrary{shapes}
\usetikzlibrary{external}

\newenvironment{packed_itemize}{
\begin{itemize}
  \setlength{\itemsep}{-0pt}
  \setlength{\parskip}{-0pt}
  \setlength{\parsep}{-0pt}
}{\end{itemize}}

\definecolor{LGrey}{gray}{0.95}

\declareuser{bc}{bc}{blue}

\graphicspath{{./fig/}}

\begin{document}

\title{Mashup of Meta-Languages and its Implementation in the Kermeta Language Workbench}

\author{Jean-Marc J{\'e}z{\'e}quel, Benoit Combemale, Olivier Barais, Martin Monperrus, Fran\c cois Fouquet
\thanks{The authors gracefully thank all present and past members of the Triskell team at IRISA for their contribution to the ideas and tools behind the Kermeta language workbench. A special thanks to Andr{\'e} Fonseca (former Master student in the team) who worked on the Kermeta-based implementation of fUML. }}

\maketitle

\begin{textblock*}{\textwidth}(0cm,-1.8cm)
\begin{center}
Accepted for publication in ``Software and Systems Modeling'' (SoSym), 2013.
\end{center}
\end{textblock*}

\begin{abstract}
With the growing use of domain-specific languages (DSL) in industry, DSL design and implementation goes far beyond an activity for a few experts only and becomes a challenging task for thousands of software engineers.
DSL implementation indeed requires engineers to care for various concerns, from abstract syntax, static semantics, behavioral semantics, to extra-functional issues such as run-time performance.
This paper presents an approach that uses one \emph{meta-language} per language implementation concern.
We show that the usage and combination of those meta-languages is simple and intuitive enough to deserve the term \emph{mashup}.
We evaluate the approach by completely implementing the non trivial fUML modeling language, a semantically sound and executable subset of the Unified Modeling Language (UML). 
\end{abstract}

\section{Introduction}

Model-driven engineering (MDE) of software fosters the use of multiple domain-specific languages (DSL) during software design and development, up to several DSLs per engineering concern and application domain \cite{schmidt06}.
In the model-driven terminology, DSLs are generally defined using metamodels and consequently, DSL programs are generally referred to as models.
For instance, a model-driven development process may use a DSL to simulate the system in prototyping phases,
another DSL to define its software architecture, and yet another one to specify the set of valid inputs so as to allow test-case generation.
The goal of using multiple DSLs is to improve various aspects of software: such as improving consistency with requirements, reducing development costs or reducing the number of bugs  \cite{schmidt06}.

However, model-driven development is no silver bullet.
One of its drawbacks is that industry, instead of relying on a small number of general-purpose languages, now
needs many modeling or programming environments of production-level quality.
In other terms, language design and implementation goes far beyond an activity for a few experts only and becomes a challenging task for thousands of software engineers and domain experts, which we call \emph{DSL engineers} \cite{Hutchinson:2011}.

DSL engineers who are responsible for designing and implementing a tool-supported DSL can of course use general-purpose programming languages such as Java.
However, implementing a DSL runtime (whether compiler of interpreter) is complex.
It requires orchestrating various concerns, as different as the definitions of abstract and concrete syntaxes, static semantics (including the well-formedness rules), behavioral semantics, as well as extra-functional issues such as compile-time or run-time performances, memory footprint, etc.

That's the reason for which researchers and innovators invent \emph{language workbenches} \cite{Ghosh:2010,fowler:2010,mernick:2013,books/daglib/0030751}. 
A language workbench provides DSL engineers with languages, libraries or tools to ease the design and implementation of DSLs.
Centaur \cite {borras1989centaur} is an early contribution in this field, more recent approaches include Metacase's MetaEdit+~\cite{tolvanen2003metaedit}, Microsoft's DSL Tools \cite{cook2007domain}, Clark et al.'s Xactium \cite{Clarka08}, Krahn et al's Monticore \cite{Krahn2008} Kats and Visser's Spoofax \cite{Kats2010} and Jetbrain's MPS \cite{VoelterSolomatov2010}.

In this paper, we present the Kermeta language workbench. In a nutshell, the Kermeta workbench involves one different meta-language per DSL implementation concern: one meta-language for the abstract syntax (aligned with EMOF \cite{omgmof2core}); one for the static semantics (aligned with OCL \cite{omgocl2}) and one for the behavioral semantics (Kermeta Language)\footnote{The concrete syntax is achieved thanks to a full compatibility with all EMF-based tools for concrete syntax, such as the de facto standards EMFText (see \url{http://www.emftext.org}) and XText (see \url{http://www.eclipse.org/Xtext/})}.
The Kermeta workbench uses an original modular compilation scheme to compose the three different meta-languages.
We do not aim at presenting  \emph{how to} define a DSL in Kermeta. This is the goal of our tutorial paper published at GTTSE \cite{DBLP:conf/gttse/JezequelBF09}. 
This paper is meant to describe the rationales and the evaluation of the main design choices of the workbench, as well as a thorough presentation of its novel compilation scheme. 

Let us now say a few words about the composition of the meta-languages in the Kermeta workbench.
The DSL concerns are organized into loosely coupled modules.
They are all compiled into traits of the Scala programming language \cite{LAMP-REPORT-2006-001}. The composition semantics we use is inspired from the concept of aspects~\cite{KiczalesLMMLLI97} and open classes~\cite{Clifton00multijava:modular}.
In practice, this combination of meta-languages only uses two keywords (``aspect'' and ``require'') and is simple and intuitive enough to deserve the term \emph{mashup}.
Implementing DSLs with the Kermeta workbench relieves DSL engineers from the burden of expressing and composing the abstract syntax together with the static and behavioral semantics.

The rest of the paper reads as follows:
\begin{itemize}
  \item Section \ref{sec:fuml-prez} describes fUML, a real world modeling DSL defined by OMG. fUML is an executable subset of the Unified Modeling Language (UML). This DSL is used throughout the paper to illustrate the concepts as well as for evaluation.
  
  \item Section \ref{sec:mahsup} presents the three meta-languages used in the Kermeta language workbench, the rationales and a qualitative evaluation of their conjunct usage to design DSLs. For a user-manual on using them, we refer the reader to~\cite{DBLP:conf/gttse/JezequelBF09}.

  \item Section \ref{sec:thecompilation} exposes the compilation scheme used to compose the three meta-languages aforementioned. In particular, the composition are described both formally and with respect to language operators of the Scala programming language. The evaluation of this section shows that the ease of design is not traded for efficiency. Indeed, the Kermeta version of fUML is as fast as the  reference Java implementation.
  
  \item Section \ref{sec:rw} compares the Kermeta workbench against the related work. The main strength of Kermeta Workbench is the use of standardized meta-languages which open doors to inter-interoperability with other tools and reusability of specifications.

\end{itemize}

Section~\ref{sec:conclusion} concludes and proposes directions of future work.

\section{The Case of fUML}
\label{sec:fuml-prez}

We consider throughout the article the case of fUML (\emph{Foundational Subset for Executable UML Models})~\cite{fuml-spec}, an executable subset of the Unified Modeling Language (UML) that can be used to define the structural and behavioral semantics of software systems.  It is computationally complete by specifying a full behavioral semantics for activity diagrams.
This means that this DSL enables implementors to execute well-formed fUML models  (here \emph{execute} means to actually \emph{run a program}). 

The main rationales of choosing fUML are that: 1) fUML is a real DSL with a non-trivial semantics, and 2) fUML has a Java reference implementation. We will compare our Kermeta design and implementation against this reference design and implementation in terms of modularity and performance.

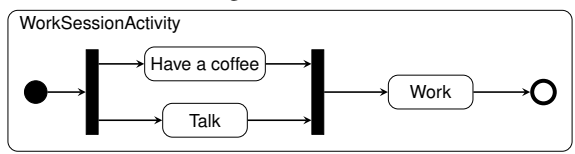
\begin{figure}[h]
  \tikzstyle{activity model}=[shape=rectangle, rounded corners=5pt, minimum width=10cm, minimum height=2.5cm, inner sep=1.6ex, draw=black]
  \tikzstyle{init node}=[shape=circle,minimum width=4mm, inner sep=0pt,fill=black, draw=black]
  \tikzstyle{fork node}=[shape=rectangle, minimum width=2mm, minimum height=15mm, fill=black, draw=black]
  \tikzstyle{join node}=[shape=rectangle, minimum width=2mm, minimum height=15mm, fill=black, draw=black]
  \tikzstyle{activity node}=[shape=rectangle, rounded corners=5pt, minimum width=15mm, minimum height=6mm, draw=black]
  \tikzstyle{final node}=[shape=circle,minimum width=4mm, inner sep=0pt,fill=none, draw=black, line width=2pt]
  \centering
  \vspace{-2ex}
  \scalebox{0.75}{
  \begin{tikzpicture}[>=stealth]
    \sffamily
    \node[activity model] (activity) {};
	\node[anchor=west] (activityName) at ($(activity.north west)+(+0.1,-0.25)$) {WorkSessionActivity};	
	\node[init node] at ($(activity.west)+(0.5,-0.2)$) (init) {};
	\node[fork node] at ($(init)+(1,0)$) (fork) {};
	\node[activity node] at ($(fork)+(2,.5)$) (coffee) {Have a coffee} ;
	\node[activity node] at ($(fork)+(2,-.5)$) (talk) {Talk} ;
	\node[join node] at ($(coffee)+(2,-.5)$) (join) {};
	\node[activity node] at ($(join)+(2,0)$) (work) {Work} ;
	\node[final node] at ($(work)+(2,0)$) (final) {};
	\draw[thick,->] (init) -- (fork) ;
	\draw[thick,->] ($(fork.east)+(0,5mm)$) -- (coffee) ;
	\draw[thick,->] ($(fork.east)+(0,-5mm)$) -- (talk) ;
	\draw[thick,->] (coffee) -- ($(join.west)+(0,5mm)$) ;
	\draw[thick,->] (talk) -- ($(join.west)+(0,-5mm)$) ;
	\draw[thick,->] (join) -- (work) ;
	\draw[thick,->] (work) -- (final) ;
  \end{tikzpicture}
}  
  \caption{Activity of the members of our team during our work sessions.}
  \label{fig:MorningStuff}
\end{figure}

\subsection{The Specification of fUML}
\label{sec:fuml}

fUML is an executable language for activity models.
As an example, Figure \ref{fig:MorningStuff} shows an executable fUML model representing the activity of our team when we meet for work sessions. We are used to first having a coffee while talking together about the latest news. When we finish to drink our coffee and to talk, we begin to work.

\paragraph{The fUML Abstract Syntax}

As illustrated by our example, the core concept of fUML is \emph{Activity} that defines a particular behavior. 
An \emph{Activity} is composed of different elements called \emph{Activity Nodes} linked by \emph{Activity Edges}.
The main nodes which represent the executable units are the \emph{Executable Nodes}. For instance, \emph{Actions} are specific  \emph{Executable Nodes} that are associated to a specific executable semantics. Other elements define the activity execution flow, which can be either a control flow (\emph{Control Nodes} linked by \emph{Control Flow}) or a data flow (\emph{Object Nodes} linked by \emph{Object Flow}). 

The example on Figure~\ref{fig:MorningStuff} uses an illustrative set of elements of the abstract syntax of fUML. The start of the \emph{Activity} is modeled using an \emph{Initial Node}. A \emph{Fork Node} splits the control flow in two parallel branches: one for the \emph{Action} of having a coffee, the other for the \emph{Action} of talking to each other. Then a \emph{Join Node} connects the two parallel branches to the \emph{Action} of working.

We refer the reader to the specification \cite{fuml-spec} for all the details about the comparison with UML2 and the whole description of the fUML abstract syntax.

\paragraph{The fUML Static Semantics}

In the specification, additional constraints are defined in order to precise the semantics and to clarify ambiguities.  fUML uses the \emph{Object Constraint Language} (OCL) in order to define those constraints. As an example, the additional constraint \emph{fUML\_is\_class}, applied over the specific action \emph{CreateObjectAction}, tells us that it can only be linked to an instance of \emph{Class} (i.e. one can not create, say activities at run-time).
In the specification, the OCL constraints are clearly separated from the rest of text.

\paragraph{The fUML Behavioral Semantics}

To support the execution of models, fUML introduces a dedicated \emph{Execution Model}. The activity execution model has a structure largely parallel to the abstract syntax using the \emph{Visitor} design pattern \cite{Gamma:1995:DPE} (called \emph{SemanticVisitor}).
Note that although the semantics is explained using visitors, which are rather at the implementation level, it is left open by the fUML specification to implement the language using other means that visitors.

In addition, to capture the behavioral semantics, the interpretation needs to  define how the execution of the activity proceeds over time. Thus, concepts are  introduced in the execution model for which there is no explicit syntax. Such concepts support the behavior of an activity in terms of \emph{tokens} that may be held by nodes and \emph{offers} made between nodes for the movement of these tokens.

Based on the execution model, the specification denotationally describes the behavioral semantics of fUML using axioms in first order logic.

\subsection{The Java Reference Implementation of fUML}
\label{subsec:fUMLReference}

ModelDriven.org\footnote{see \url{http://portal.modeldriven.org/}} is a consortium of government, commercial and university members which develops model-driven technologies.
One of the consortium projects is to provide a reference implementation of fUML\footnote{see \url{http://portal.modeldriven.org/project/foundationalUML}}.
This implementation is written in Java and is capable of executing any valid fUML model which are specified in an XML-based format. 
In this reference implementation, all the concerns of the DSL described above are written in Java.

\section{The Specification and Design of DSLs in Kermeta}
\label{sec:mahsup}

Kermeta is a language workbench designed for specifying and designing DSLs.
For this, it involves different\\meta-languages depending on the concern: abstract syntax (we will also use the term ``metamodel'' to refer to it\footnote{This is one definition in the community. For some researchers, ``metamodel'' sometimes referred to abstract syntax plus static semantics.}), static semantics, behavioral semantics. 

The workbench integrates the OMG standards EMOF and OCL, respectively for specifying the abstract syntax (cf. Subsection \ref{subsec:mof}) and the static semantics (cf. Subsection \ref{subsec:ocl}). 
The choice of EMOF and OCL has been driven by the fact that they are de facto standards. This allows full interoperability with other tools. Indeed, during the years of development of Kermeta, we have reused EMOF models and OCL constraints from many different sources.
The workbench also provides \emph{Kermeta Language} to address the specification of the operational semantics (cf. Subsection \ref{subsec:kermeta}).

The Kermeta Workbench also provides composition operators responsible for mashing-up these different concerns into a standalone execution engine (interpreter or compiler) of the DSL (cf. Subsection \ref{subsec:aspectoriented}).

In this section we illustrate all these features by presenting the implementation of fUML. Then we evaluate the use of the Kermeta workbench from the end user point of view (here, the DSL engineer) on the basis of this case study (cf. Subsection \ref{subsec:eval-user}).

\subsection{Concern \#1: Abstract Syntax Definition}
\label{subsec:mof}

First of all, to build a DSL in Kermeta, one defines its abstract syntax (i.e., the metamodel), which specifies the domain concepts and their relations.
The abstract syntax is expressed in an object-oriented manner, using the OMG meta-language EMOF (Essential Meta Object Facility) \cite{omgmof2core}.
EMOF provides the following language constructs for specifying a DSL metamodel:
package, classes, properties, multiple inheritance and different kinds of associations between classes.
The semantics of these core object-oriented constructs is close to a standard object model that is shared by various languages (\emph{e.g.,} Java, C\#, Eiffel). 
\emph{We chose EMOF for the abstract syntax because it is a \emph{de facto} standard allowing interoperability with other tools}.

\begin{figure*}
  \centering
  \tikzumlset{font=\sffamily,fill class=white}
  \scalebox{0.75}{
	\begin{tikzpicture}[x=2.8cm, y=2.5cm]
	  \umlclass[y=0, x=-0.3]{Activity}{
		name : EString
	  }{}
	  \umlclass[y=-1, x=-1.5,type={\,abstract\,}]{ActivityNode}{
		name : EString
	  }{}
	  \umlclass[y=-1, x=1,type={\,abstract\,}]{ActivityEdge}{}{}
	  
	  \umlclass[y=-2, x=-2.5]{ObjectNode}{}{}
	  \umlinherit[geometry=|-|]{ObjectNode}{ActivityNode}
  
	  \umlclass[y=-2, x=-1.5,type={\,abstract\,}]{ControlNode}{}{}
	  \umlinherit[geometry=|-|]{ControlNode}{ActivityNode}
	  
	  \umlclass[y=-2, x=-0.45,type={\,abstract\,}]{ExecutableNode}{}{}
	  \umlinherit[geometry=|-|]{ExecutableNode}{ActivityNode}
	
	  \umlclass[y=-2, x=0.55]{ControlFlow}{}{}
	  \umlinherit[geometry=|-|]{ControlFlow}{ActivityEdge}
	  \umlclass[y=-2, x=1.45]{ObjectFlow}{}{}
	  \umlinherit[geometry=|-|]{ObjectFlow}{ActivityEdge}
	
	  \umlclass[y=-3, x=-3]{InitialNode}{}{}
	  \umlinherit[geometry=|-|]{InitialNode}{ControlNode}
	  \umlclass[y=-3, x=-2]{ForkNode}{}{}
	  \umlinherit[geometry=|-|]{ForkNode}{ControlNode}
	  
	  \umlclass[y=-2.9, x=-0.45,type={\,abstract\,}]{Action}{}{}
	  \umlinherit{Action}{ExecutableNode}
	
	  \umlclass[y=-3.7, x=-2.5]{FinalNode}{}{}
	  \umlinherit[geometry=|-|,arm1=1.2]{FinalNode}{ControlNode}
	  \umlclass[y=-3.7, x=-1.5]{JoinNode}{}{}
	  \umlinherit{JoinNode}{ControlNode}
	
	  \umlclass[y=-3.7, x=-0.45]{CreateObjectAction}{}{}
	  \umlinherit{CreateObjectAction}{Action}
	  
	  \umlunicompo[mult=node\quad 0..*, pos=0.8]{Activity}{ActivityNode}
	  \umlunicompo[arg=0..*\quad edge, pos=0.8]{Activity}{ActivityEdge}
	  \umlCNrelation[
		anchor1=20, mult1=source,arg1=1,pos1=0,align1=left,
		anchor2=160,mult2=outgoing,arg2=0..*,pos2=2.0,align2=right
	  ]{ActivityNode}{-0.3,-0.75}{ActivityEdge}
	  \umlCNrelation[
		anchor1=-20,mult1=1,arg1=target,align1=left,pos1=0,
		anchor2=-160, mult2=0..*,arg2=incoming,pos2=2.0,align2=right
	  ]{ActivityNode}{-0.3,-1.25}{ActivityEdge}
	\end{tikzpicture}
  }
  \caption{Excerpt of the fUML Metamodel}
  \label{fig:fumlMetamodel}
\end{figure*}
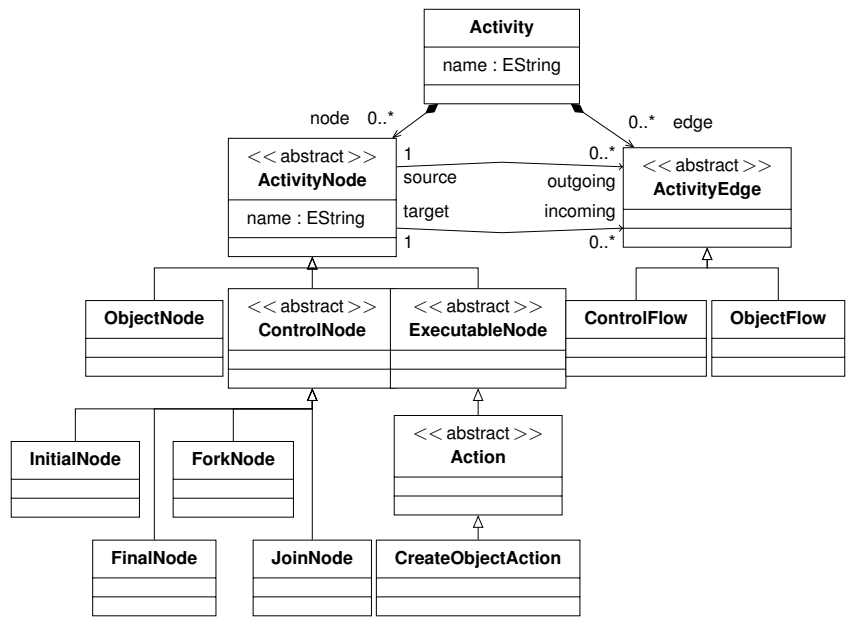

Figure \ref{fig:fumlMetamodel} shows the excerpt of the fUML metamodel explained in the previous section, and depicts those classes as a class diagram. 
In our Kermeta-based fUML design, we reuse the abstract syntax standardized by the OMG. In practice, OMG provides the fUML metamodel in terms of EMOF and we automatically translate it into an Ecore-based metamodel (the format supported by the Kermeta workbench).

Since the abstract syntax is expressed as an object-oriented metamodel, a concrete fUML model (equivalent to a DSL progam) is composed of instances of the metamodel classes.

\subsection{Concern \#2: Static Semantics Definition}
\label{subsec:ocl}

The static semantics of a DSL is the union of the well-formed rules on top of the abstract syntax (as invariants of domain classes) and the axiomatic semantics (as pre- and post conditions on operations of metamodel classes).
The static semantics is used to statically filter incorrect DSL programs before actually running them.
It is also used to check parts of the correctness of a DSL program's execution either at design-time using model-checking or theorem proving, or at run-time using assertions, depending on the execution domain of the DSL. Kermeta uses the OMG Object Constraint Language (OCL) \cite{omgocl2} to express the static semantics, directly woven into the metamodel using the Kermeta aspect keyword.

Listing \ref{lst:ocl-fuml} shows the previously introduced fUML invariant (specified with the keyword ``inv'') as expressed in the Kermeta workbench using OCL.

\begin{lstlisting}[language=kermeta,
    caption={Weaving the Static Semantics of fUML into the Standard Metamodel},
    label={lst:ocl-fuml}, 
    basicstyle=\small\ttfamily, backgroundcolor=\color{LGrey}, numbers=left, xleftmargin=20pt]
package fuml;
require "fuml.ecore"
aspect class CreateObjectAction {
// The given classifier must be a class.
inv fUML_is_class :
  self.classifier.oclIsKindOf(Class)
}
\end{lstlisting}

In the Kermeta workbench, the abstract syntax and the static semantics are conceptually and physically (at the file level) defined in two different modules.
Consequently, it is possible to define different semantic variants for the same domain, i.e. to have a single EMOF metamodel shared by different static semantics, \emph{e.g.,} to cope with language variants.

\subsection{Concern \#3: Behavioral Semantics Definition}
\label{subsec:kermeta}

EMOF does not include concepts for the definition of the behavioral semantics and OCL is a side-effect free language.
To define the behavioral semantics of a DSL, we have created the Kermeta Language, an action language that is used to express the behavioral semantics of a DSL~\cite{models05pam}.
It can be used to define either a translational semantics or an operational semantics~\cite{jsw09}. A translational semantics would result in a compiler while an operational semantics would result in an interpreter. In this paper, for sake of clarity, we will only present operational semantics. However, the language is exactly the same in both cases.

Using the Kermeta language, an operational semantics is expressed as methods of the classes of the abstract syntax~\cite{models05pam}. Listing~\ref{lst:OpSemantics} is an excerpt of the operational semantics of fUML. Using the \texttt{aspect} keyword, a method ``execute'' is added to the metamodel class ``Activity''. The body of the method imperatively describes what is the effect of executing an activity. In this case, it consists of i) creating an activity node activation group and activating all the activity nodes in the activity, and ii) copying the values on the tokens offered by the output parameter nodes to the corresponding output parameter. 

The Kermeta language is imperative, statically typed, and includes classical control structures such as blocks, conditionals, loops and exceptions. The Kermeta language also implements traditional object-oriented mechanisms for handling multiple inheritance and generics.  The Kermeta language provides an execution semantics to all EMOF constructs that must have a semantics at run-time such as containment and associations. First, if a reference is part of a bidirectional association, the assignment operator semantics handles both ends of the association at the same time.
Second, if a reference is part of a containment association, the assignment operator semantics unbinds existing references if any, so that one object is part of another one. Finally, for multiple inheritance, Kermeta borrows the semantics from the Eiffel programming language~\cite{Meyer:1992:EL:129093}.

\begin{lstlisting}[language=kermeta,
caption={In Kermeta, the DSL operational semantics is expressed as methods of metamodel classes},float,
label={lst:OpSemantics}, 
basicstyle=\small\ttfamily, backgroundcolor=\color{LGrey}, numbers=left, xleftmargin=20pt]
aspect class Activity inherits Executable{
// the semantics of executing an activity
method execute(runnable : Runnable) : Void is do					
	// Creation of an activity node activation group
	runnable.execute()
	var group : ActivityNodeActivationGroup init ActivityNodeActivationGroup
				.new()
	group.execution := runnable
	runnable.group := group

	// Activation of all the activity nodes in the activity
	runnable.group.activate(self.node, 
		self.edge)			
	var outputNodeActivations : 
		OrderedSet<ActivityParameterNode> 
		init runnable.group.
	fumlGetOutputParameterNodeActivations()

	// Copy the values on the tokens offered by output parameter nodes to the corresponding output parameters
	outputNodeActivations.each {outputNodeActivation |
		var parameterValue : ParameterValue init ParameterValue.new()
		parameterValue.parameter := (outputNodeActivation.asType(ActivityParameterNode)).parameter
		var tokens : Set<Token> init outputNodeActivation.
					fumlGetTokens()
		tokens.each { token |
			var val : Value init (token.asType(ObjectToken)).val
			if (val != void) then
				parameterValue.values.add(val)
			end
		}			
		runnable.fumlSetParameterValue(parameterValue)
	}
end
}
\end{lstlisting}

\subsection{Composition Operators for the Mashup of Meta-Languages}
\label{subsec:aspectoriented}

As introduced above, mashing-up all DSL concerns in the Kermeta workbench is achieved through two keywords in the Kermeta language: \texttt{aspect} and \texttt{require}.

In Kermeta, all pieces of static and behavioral semantics are encapsulated in metamodel classes.
For instance, in Listing \ref{lst:OpSemantics}, the behavioral semantics  are expressed in the metamodel classes ``Activity''.
The \texttt{aspect} keyword enables DSL engineers to relate the language concerns (abstract syntax, static semantics, behavioral semantics) together.
It allows DSL engineers to reopen a previously created class to add some new pieces of information such as new methods, new properties or new constraints. It is inspired from open-classes \cite{Clifton00multijava:modular} and will be further discussed in Section \ref{sec:thecompilation}.

The keyword \texttt{require} enables the mashup itself. 
A DSL implementation \emph{requires} an abstract syntax, a static semantics and a behavioral semantics.
Listing \ref{lst:fumlmashup} shows how the final fUML mashup looks like in Kermeta.
Three \texttt{require} are used to specify the three concerns.
The require mechanism also provides some flexibility with respect to static and behavioral semantics. For example, several behavioral semantics could be defined in different modules (all on top on the same metamodel) and then chosen depending on particular needs (e.g., simulation, compilation).
It is also convenient to support semantic variants of a given language. For instance, Kermeta can be used to specify several implementations of UML semantics variation points.

\begin{lstlisting}[language=kermeta,
    caption={Mashup of the fUML Concerns},
    label={lst:fumlmashup}, 
    float,basicstyle=\small\ttfamily, backgroundcolor=\color{LGrey}, numbers=left, xleftmargin=20pt]
package fuml;
require "fuml.ecore" // abstract syntax
require "fuml.ocl"  // static semantics
require "fuml.kmt"  // operational semantics
class Main {
  operation Main(): Void is do
    // Execute a fUML model
  end
}
\end{lstlisting}

\subsection{Evaluation of the Mashup-based Design From the End-user Viewpoint}
\label{subsec:eval-user}

The Kermeta-based design of fUML follows the approach described in this section. It means that all fUML concerns (abstract syntax, statics semantics and behavioral semantics) are separated in different units and that the fUML runtime environment is the result of the mashup. In other terms, the Kermeta-based design clearly separates the three concerns of DSL implementation, and thus loyally reflects the structure of the specification.

For us, the driving motivation of having three different modules for implementing a DSL  is:
1) allowing that each part of a DSL is done by different stakeholders (e.g. the abstract syntax by the standardization body and the compiler by the tool vendors), possibly in parallel;
2) supporting different variation points (either syntactic or semantic).
The main con is that there is one DSL engineer, responsible for the meta-language integration, who must understand the different modules.

Jezequel et al. \cite{DBLP:conf/gttse/JezequelBF09} presented a reference manual on how to design and implement a DSL with the Kermeta language workbench. Beyond \emph{how to}, we present in this section the advantages of our approach from the viewpoint of the DSL engineer (the end-user in our context).

\subsubsection{Are concerns designed in different modules?}
\label{claim:different-modules}
Our mashup approach provides two dimensions of modularity: 
\begin{itemize}
\item modularity of domain concepts (metamodel concepts) 
\item modularity of language engineering concerns (parsing, static semantics, interpretation, compilation, etc.)
\end{itemize}

The first one of course does not bring anything new with respect to a Java
based approach (e.g. the fUML reference implementation): this is just the
usual class-based modularity found in OO languages, including MOF. Still it is
very helpful when one wants to slightly change an existing concept, or even add
or remove one into the meta-model.

So the originality of our approach lies in the modularity of language engineering concerns.    
It is accepted that the design of a modeling language deals at least with~\cite{Harel2004}: the abstract syntax  (called a metamodel in the model-driven engineering terminology), the set of constraints on the abstract syntax (called static semantics), and the behavioral semantics.
Using our approach, all these concerns are implemented in different modules:
\begin{itemize}
\item the abstract syntax is defined as an EMOF metamodel~\cite{omgmof2core}. Technically speaking, it is completely defined in an Ecore file (e.g. \texttt{fuml-metamodel.ecore}). 
Please refer to~\cite{emf2} for more details about this file format. 
This module is standalone and does not depend from others modules.
\item the static semantics is defined in a module dedicated to invariants, pre- and postconditions. When using OCL, this means creating a file, say \texttt{fuml.ocl}. This module imports the language metamodel (the Ecore file aforementioned) and has no other dependencies.
\item the behavioral semantics (also known as execution semantics) is defined in a dedicated module using the Kermeta language, say \texttt{fuml.kmt}.
\end{itemize}

There is an exact one-to-one mapping between the abstract concerns of language design and the concrete design modules. The design of the modules are clearly layered so that there are no more spurious design dependencies than logical dependencies.

This architecture also enables DSL engineers to get rid of certain heavyweight design patterns, such as the design pattern \emph{Visitor} which is often used to inject the semantics.
Not using such patterns has two advantages. First, the DSL design is easier to understand and maintain.
Second, at run-time, our DSL architecture requires less communication between objects (delegates and proxy calls), which contributes to a better efficiency (this point is further discussed in \ref{sec:thecompilation}).

\subsubsection{Are concerns designed using appropriate\\ meta-languages?}
The research on aspect-oriented software development has shown \cite{shonle2003xaspects} that not all languages are equals to implement aspects. Certain concerns are well-suited to be implemented using domain-specific languages, they are sometimes called domain-specific aspect languages.

Using our approach, all concerns are designed using \emph{meta-}DSLs.
The abstract syntax is designed using EMOF, the static semantics is designed using OCL and the operational semantics is designed using the Kermeta language.
Let us now review the advantages of certain particular \emph{meta-}DSLs for the related concerns.

\paragraph{Metamodeling with EMOF}
\label{metamodeling-emof}

We argue that EMOF is really appropriate to design DSL metamodels:
\begin{itemize}
\item it is based on object-oriented modeling. Thus, any engineer who is fluent with object-oriented thinking is able to intuitively design a language metamodel with EMOF.
\item it is the result of years of discussion between DSL experts: it contains a lot of constructs that are known to be useful for metamodeling (\emph{e.g.,} association containment and multiple inheritance).
\item the tool-support for EMOF is good: there are several vendors and mature tools; it is possible to express EMOF metamodels using different textual and graphical editors.
\end{itemize}

\paragraph{Static Semantics with OCL}

OCL is well-suited for defining a DSL static semantics for the same reasons as those presented in the previous paragraph (maturity and tool support). Furthermore, since OCL is side-effect free, it is impossible for DSL engineers to accidentally introduce some pieces of behavioral semantics in the static semantics definition. In other words, the DSL design itself participates to ensuring the separation of concerns. On the contrary, using Java/AspectJ or another general-purpose programming language for expressing the static semantics would open the door to concern tangling.

\paragraph{Operational Semantics with the Kermeta Language}

As already stated, Kermeta is a workbench as well as a language.
As a language (\emph{Kermeta Language}), it has been specifically designed to express the operational semantics of languages.
Let us now review a couple of examples that show the power of the Kermeta language with this respect.

\noindent\emph{Manipulating collections of objects: }
The operational semantics of DSLs often deals with manipulating collections of objects. For instance a DSL for state machines would at some point traverse all transitions starting from a given state.
Kermeta includes useful functions based on lambda expressions to manipulate collections: e.g. \emph{collect}, \emph{select} or \emph{reject} (similar functions can be found in OCL and in some general purpose languages such as Smalltalk). Those constructions give a natural way to navigate through models compared to iteration over Java collections.

\noindent\emph{Manipulating metamodel concepts within the operational semantics: }
When implementing a behavioral semantics, one often manipulates concepts of the metamodel.
If the language used to implement operational semantics does not natively support metamodeling concepts,
the operational semantics is bloated with workarounds to approximate the metamodeling concepts.
For example, Figure \ref{fig:pinModel} represents an excerpt of the fUML implementation where the \emph{Pin} class inherits from both \emph{ObjectNode} and \emph{MutiplicityElement}, and \emph{isReady} is a method of \emph{Pin} that expresses a piece of operational semantics. Let us compare how to handle two excerpts of this method in Java and Kermeta.

This listing emphasizes one important characteristics of our approach.
The Kermeta language enables to directly manipulate the concepts of the language without having to use special wrapper methods.
For instance, on the right-hand side listing, \texttt{lower} directly refers to the field \emph{lower} of the metamodel class \emph{MutiplicityElement}. On the contrary, in Java, the DSL engineer always has to master the simulation of the semantics of multiple inheritance, association, containment, etc. In other words, our approach lowers the representational gap between the code of the operational semantics and the metamodel concepts (\emph{i.e.} of the abstract syntax).
Also, writing the operational semantics with the Kermeta language enables to avoid the bloating (code generation, annotations, etc.) due to embedding the powerful semantics of the metamodeling language (EMOF) into a programming language that does not support it by default (\emph{e.g.,} Java).

\begin{figure*}
\begin{minipage}{9cm}
\begin{lstlisting}[language=Java,
    basicstyle=\scriptsize\ttfamily, backgroundcolor=\color{LGrey}]
/****** In Java *********/
class Pin extends ObjectNode {
  // simulates multiple inheritance using delegation
  public multiplicityElement = new MultiplicityElement();
}
// elsewhere
class InputPinActivation extends PinActivation {
  public boolean isReady() {
    // The DSL engineer has to know
    // all low-level implementation choices, here the 
    // use of delegation to simulation multiple inheritance
    int minimum = this.node.multiplicityElement.lower;
  }
}
\end{lstlisting}
\end{minipage}
\begin{minipage}{9cm}
\begin{lstlisting}[language=kermeta,
    basicstyle=\scriptsize\ttfamily, backgroundcolor=\color{LGrey}]
/****** In Kermeta *********/

aspect class Pin inherits ObjectNode, MutiplicityElement {
 ... 
}

aspect class InputPinActivation {

  operation isReady() : Boolean is do
    // "lower" from MultiplicityElement on node
    var minimum : Integer init self.node.lower
  end

}
\end{lstlisting}
\end{minipage}
\caption{Comparison of the Java Implementation against the Kermeta Implementation. The programmer is relieved from the accidental complexity due to the code generation.}
\label{fig:pinModel}
\end{figure*}

\section{The Compilation of the Mashup of Meta-Languages in Kermeta}
\label{sec:thecompilation}

This section presents how we compile a mashup of three meta-languages into a single DSL. This poses a number of challenges, described in Subsection~\ref{sec:challenges} that we solve by generating Scala code as exposed in Subsection~\ref{sec:whyscala}. The resulting code satisfies the composition semantics expressed at the meta-language level (cf. Subsection~\ref{sec:compilation}).

\subsection{Challenges}
\label{sec:challenges}

We have to choose a single appropriate target language for compiling the mashup.  As input, we have three different meta-languages (Ecore, OCL, Kermeta). For sake of simplicity, the target language must not be too far from the concepts and the abstraction level of our meta-languages.
While many modern programming language can be appropriate (e.g. Java, C\#, Python), we have identified the following challenges that should be tackled in our context:

\noindent\textbf{Challenge \#1: Expressing the composition semantics}
The target language must not only be appropriate for supporting the compilation of each
meta-language, it must also support the expression of our composition semantics (\emph{\`a la} open-class) used in the Kermeta language.

\noindent\textbf{Challenge \#2: Integration with legacy code} 
We described in the introduction that DSL engineers have strong incentives to design their language in a way that is interoperable with other language engineering tools.
For instance, there is today a strong ecosystem of language engineering tools around the EMF platform.
The core of EMF is a generator that outputs Java code responsible for handling the metamodeling semantics and marshaling (read and save DSL programs from disk).
The second design challenge of our compilation chain is to plug our mashup compiler onto the untouched code generated from EMF.

\noindent\textbf{Challenge \#3: Efficiency} 
The approach aims to improve the design of DSLs in providing support of separation of concerns; one of the risk is to introduce a performance overhead.  The last challenge is to produce executables as efficient as an ad-hoc implementation.

\subsection{Using Scala as a Target Language}\label{sec:whyscala}

Our mashup compiler generates Scala code from the three language concerns. Indeed, Scala is a solution to the three aforementioned challenges:
1) it has a low gap with OCL and Kermeta. In particular, OCL and Kermeta closures are straightforwardly compiled to Scala closures ;
2) Scala's mixin composition semantics is a nice building block for defining our open-class composition semantics at the meta-language level ;
3) it is able to seamlessly use the Java code generated from the EMF compiler ;
4) Scala is known to be efficiently compiled into bytecode and there has been significant work on Scala performance~\cite{citeulike:10319713}.

\subsection{Mashup Compilation Scheme}\label{sec:compilation}

\begin{figure}
	\centering
	\includegraphics[width=1.\columnwidth]{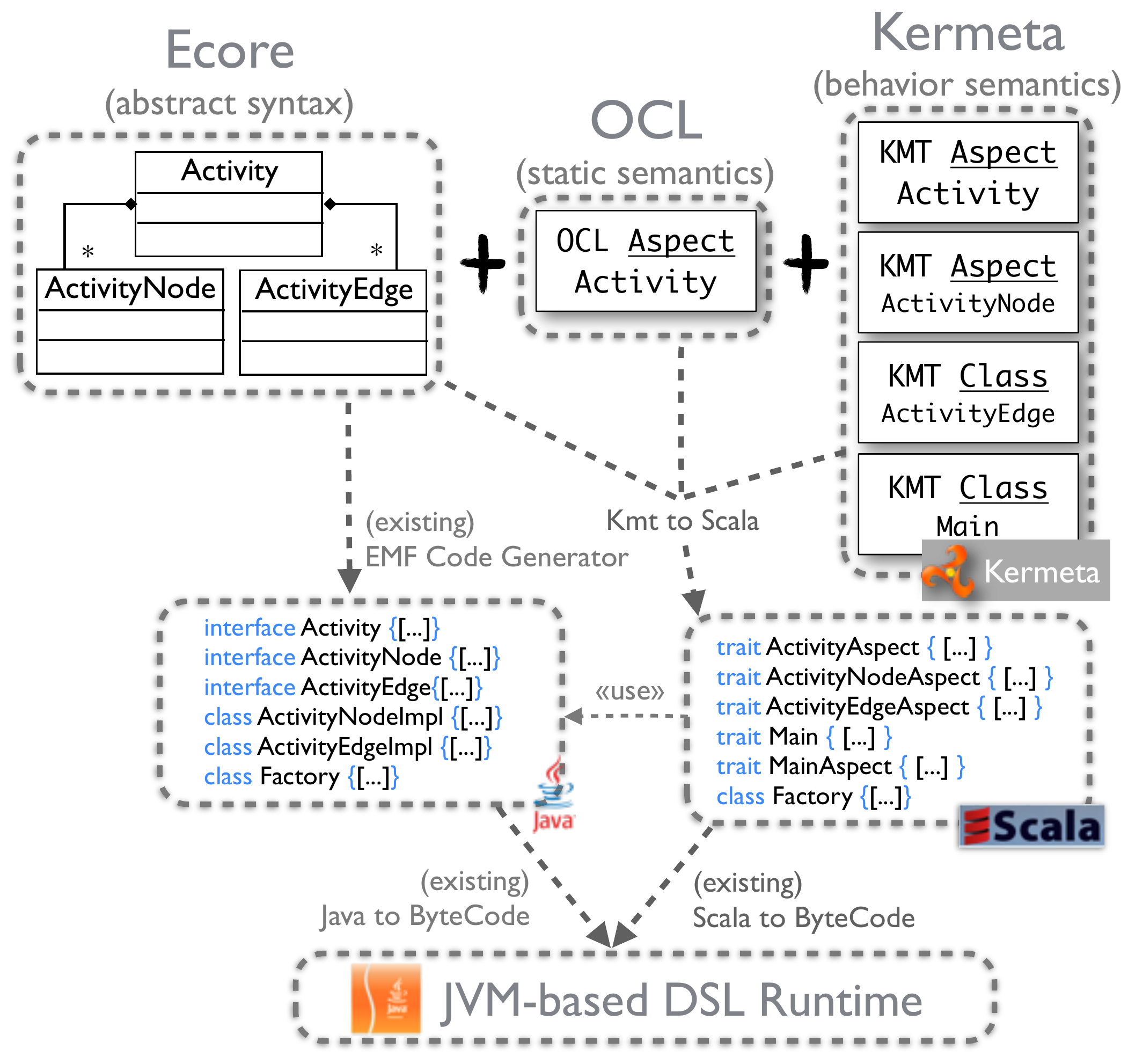} 
	\caption{Java and Scala Elements Generated from a Simple Ecore and Kermeta Metamodel}
	\label{fig:openclass}
\end{figure}

The main input of the mashup is a set of Ecore classes, the static semantics defined in OCL and the behavioral semantics defined in Kermeta (see Figure \ref{fig:openclass}). To build the mashup at the bytecode level, the mashup compiler must plug new generated code into the generated, untouched code from EMF. 

\subsubsection{Overview}

The compilation scheme relies on six main generated artifacts as follows. Most explanations refer to listing \ref{lst:scalaopenclass} which shows an excerpt of the generated Scala code.

\begin{enumerate}
\item EMF generates one Java class ($A_{Java_{C}}^{Emf}$) and one Java Interface ($A_{Java_{I}}^{Emf}$) per type of the abstract syntax (that we now call respectively metaclass and metainterface).
\item EMF generates one factory class ($Factory_{Java}^{Emf}$) per Ecore package.
\item The Kermeta compiler generates one trait per class for the mashup of the static semantics and the behavioral semantics ($A_{Scala_{T}}^{Kmt}$) (\emph{e.g.,} \texttt{ActivityAspect}, see line 9).
\item The Kermeta compiler generates one Scala class for composing the EMF class and the aforementioned traits ($RichA\-_{Scala_{C}}$) (\emph{e.g.,} \texttt{class RichActivity}, see line 23. Note that\\ \texttt{ActivityImpl} is generated by EMF).
\item The Kermeta compiler generates one Scala singleton object that extensively uses Scala implicits\\ (\emph{e.g.,} \texttt{ImplicitConversion}, see lines 37--53).
\item The Kermeta compiler generates a factory Scala object ($Rich\-Factory$ $(RichA_{Scala_{C}})_{Scala}$) that overrides the default EMF factory (\emph{e.g.,} \texttt{RichFactory}, see lines 32--36).
\end{enumerate}

\subsubsection{Compilation Scheme of the Open-Class Mechanism}

Let us now summarize the semantics of the open-class mechanism provided by Kermeta as implemented in the compiler, and relying on existing the following object oriented composition operators Java $extends$ ($\bullet$), Java $implements$ ($\circ$), Scala $extends$ ($\odot$) and Scala $with$ ($\Diamond$). Since the abstract syntax is compiled using the EMF generator, the main goal of this specification is to show how Kermeta permits a fine grained integration with the Java code generated by EMF. 

To compose two classes $A$  and $B$ ($A \oplus B$) that share the same signature using Kermeta open class operator, three cases must be considered.

\paragraph{Case 1.} $A_{Kmt}$ and $B_{Kmt}$, i.e. the two different classes are defined using the Kermeta language.  For this case, the composition is done at the metamodel level. Thus, there is no integration issue at the code level. This case can be transformed to the composition of a Kermeta $C$, that results from the composition of $A$ and $B$ at the model level.
\begin{center}
\vspace{-5ex}
 \begin{align}\label{eqn:vi}
A_{Kmt}^{base} \oplus B_{Kmt}^{aspect} = C_{Kmt}^{base} 
 \end{align}
\end{center}

\paragraph{Case 2.} $A_{Ecore}$ and $B_{Ecore}$, i.e. the two different classes are defined using Ecore and composed using the Kermeta language.  For this case, the composition is forbidden. Kermeta does not allow mixing of two Ecore classes having the same signature. 

\paragraph{Case 3.} $A_{Ecore}$ and $B_{Kmt}$ (or $A_{Kmt}$ and $B_{Ecore}$ respectively), i.e. a class defined using Ecore is composed with a class defined using the Kermeta language. This case requires a fine grained integration at the code level between the legacy EMF code and the Kermeta aspect code.  In this case $A \oplus B = A_{Ecore} \oplus B_{Kmt}$.

From $A_{Ecore}$, we obtain the following building blocks:

\vspace{-5ex}
\begin{center}
 \begin{align}\label{eqn:iii}
(i)~A_{Java_{C}}^{Emf} \bullet super(A)_{Java_{C}}^{Emf}  \circ A_{Java_I}^{Emf}\\ 
\notag (ii)~Factory(A)_{Java}^{Emf} 
 \end{align}
\end{center}

where $super$ is a function which returns all the parent classes of a given class.

From $B_{Kmt}$, we obtain the following building blocks:

\vspace{-5ex}
\begin{center}
 \begin{align}\label{eqn:iv}
B_{Scala_{T}}^{Kmt} \odot super(B)_{Scala_{T}}^{Kmt}
 \end{align}
\end{center}

Finally, the composition provides:

\vspace{-5ex}
\begin{center}
 \begin{align}\label{eqn:v}
(i)~RichA_{Scala_{C}} \bullet A_{Java_{C}}^{Emf} \odot B_{Scala_{T}}^{Kmt} \\
\notag (ii)~RichFactory(RichA_{Scala_{C}})_{Scala} \bullet Factory(A)_{Java}^{Emf}\\
\notag (iii)~ImplicitConversion~A2RichA \& B_{Scala_{T}}^{Kmt}2RichA 
\end{align}
\end{center}

The role of the \texttt{ImplicitConversion} object is discussed in section \ref{sec:implicit}.

\subsection{Specific Issues}
Let us now describe in details some points of interest of our compiler.

\subsubsection{Integration with EMF}

Our generated code integrates with EMF by extending EMF classes in the object-oriented sense. 
For instance, in listing \ref{lst:scalaopenclass}, the class \texttt{RichActivity} extends the class \texttt{ActivityImpl} which is generated by EMF.
Also, we leverage the design pattern ``abstract factory'' implemented by EMF for overriding the creation of instances. The generated factory \texttt{RichFactory} overrides the EMF factory \texttt{FactoryImpl} and returns objects that are enriched with static and behavioral semantics.
As a result, all EMF-compatible programs are interoperable with our DSL runtime.

\subsubsection{Compiling the Static Semantics}

The static semantics is a set of invariants and pre/post conditions (well-formed rules). For each metaclass, we generate a Scala trait that contains the corresponding invariants and pre and post conditions.

To preserve substitutability, the static semantics respects the Liskov substitution principle~\cite{Liskov94abehavioral}. Liskov's principle imposes some standard requirements on signatures that have been adopted in newer object-oriented programming languages.  In addition to the traditional covariance and contravariance rules, there is a number of behavioral conditions leading to some restrictions on how contracts can interact with inheritance. 
\begin{itemize}
\item Rule 1: Invariants of the supertype must be preserved in a subtype. 
\item Rule 2: Preconditions cannot be strengthened in a subtype, postconditions cannot be weakened in a subtype~\cite{Meyer:1992:EL:129093}. 
\end{itemize}

Scala provides building blocks to help support Design by Contract\footnote{\url{https://wiki.scala-lang.org/display/SYGN/Design-by-contract}}~\cite{Odersky:2010:CS:1939399.1939405}.  To satisfy Rule~1,  we choose to flatten the invariant inheritance at compile-time. Consequently each Scala trait representing a metaclass contains its own invariants and all the invariants of its super classes. Pre- and posconditions are also flattened statically.  Consequently we do a logical disjunction between the preconditions of an operation and the preconditions of its super operation, we do a logical conjunction between the postconditions of an operation and the postconditions of its super operation.

\subsubsection{Expressing the Composition}

The semantics of composition that we have introduced at the meta-language level (keyword ``aspect'') is implemented using Scala traits as follows:

\begin{lstlisting}[breaklines=true, language=scala, basicstyle=\small\ttfamily, backgroundcolor=\color{LGrey}, xleftmargin=0pt]
class Activity extends ActivityImpl with ActivityAspect
\end{lstlisting}

Indeed, Scala's class linearization mechanism\footnote{
In Scala, the classes and traits inheritance hierarchy forms a directed acyclic graph (DAG). The term linearization refers to the algorithm used to ``flatten'' this graph for the purposes of resolving method lookup priorities, constructor invocation order, binding of super, etc. The linearization defines the order in which method lookup occurs. We refer the interested reader to \cite{LAMP-REPORT-2006-001} for an exhaustive explanation of the Scala's class linearization mechanism.} mixes-in the two language implementation concerns. \texttt{ActivityImpl} comes from EMF and contains the abstract syntax definition of Activity, \texttt{ActivityAspect} is generated from the OCL specification of the static semantics and the Kermeta behavioral semantics.
For type-checking, the trait uses an implicit object behind the scene that we present in \ref{sec:implicit}.
Let us now explain how we deal with Kermeta multiple inheritance.

Since EMOF offers multiple inheritance, Kermeta has to support it. When two methods with the same signature are present, the programmer must specify an explicit renaming. With this mechanism, developers can explicitly resolve the ambiguity by choosing one of super methods.  In Scala, you can invoke the supertype using the super keyword, or you can directly reference any of the traits in the declaration of the class, possibly skipping some of the overridden methods of the traits between, by qualifying the super keyword with a trait type (\textit{super[superClassName]}). The mashup compiler uses this mechanism for choosing in which branch we want to call the super method.

\subsubsection{The Need for Scala's Implicits}\label{sec:implicit}

To support the openclass mechanism, at the JVM level, a $Activity\-Node$ class would have to be extended with the static semantics and the behavioral semantics integrated with Kermeta which is impossible without modifying the bytecode of the class.  Since only object typed as \textit{e.g.,} \texttt{Rich\-Activity\-Node} or \texttt{Activity\-Node\-Aspect} can invoke the method \texttt{fire}, an object typed as a \texttt{Activity\-Node} can not see the added method.  
In order for base instances to transparently be seen as enriched objects, the mashup compiler generates a global conversion object named \texttt{Implicit\-Conversion}. This object is imported in each class. It declares an implicit conversion method for converting base to rich class and aspect to rich class as illustrated in Listing~\ref{lst:scalaopenclass}. This conversion mechanism is based on the Scala view operator.

A view from type $S$ to type $T$ is defined by an implicit value which has function type $S \rightarrow T$, or by a method convertible to a value of that type. Views are applied in two situations: (i) If an expression $e$ is of type $T$, and $T$ does not conform to the expression's expected type $pt$. (ii) In a selection $e.m$ with $e$ of type $T$, if the selector $m$ does not denote a member of $T$. In the first case, a view $v$ is searched which is applicable to $e$ and whose result type conforms to $pt$. In the second case, a view $v$ is searched which is applicable to $e$ and whose result contains a member named $m$.

With this mechanism, Java code that uses EMF generated types can view new methods and attributes as soon as it imports the \texttt{Implicit\-Conversion}
object. 
We also reuse the view mechanism to provide an implicit conversion between EMF Collections and Scala Collections.

\subsection{Compilation Example}

\begin{lstlisting}[float,breaklines=true, language=scala, basicstyle=\small\ttfamily, backgroundcolor=\color{LGrey}, numbers=left, xleftmargin=20pt,caption=Excerpt of the generated Scala code from Figure \ref{fig:openclass}, label=lst:scalaopenclass]
import kermeta.implicitconversion._


trait Main
trait MainAspect extends k2.core.Object {
	def newOp() = { ... }
}

trait ActivityAspect extends k2.core.Object, ExecutableAspect{
	def execute() = { ... }
}

trait ActivityNodeAspect extends 
		k2.core.Object{
	def run() = { ... }
}

trait ActivityEdgeAspect extends 
		k2.core.Object{
	def run() = { ... }
}

class RichActivity extends ActivityImpl with ActivityAspect
class RichActivityNode extends ActivityNodeImpl 
		with ActivityNodeAspect
class RichActivityEdge extends ActivityEdgeImpl 
			with ActivityEdgeAspect
class RichMain extends EObjectImpl 
			with Main 
			with MainAspectAspect

object RichFactory extends FactoryImpl{
	override def createActivity : Activity = {new RichActivity }
	def createMain : Main = {new RichMain}

}
object ImplicitConversion {
	implicit def rich(v : Activity) = 
			v.asInstanceOf[RichActivity]
	implicit def rich(v : ActivityAspect) = v.asInstanceOf[RichActivity]
	implicit def rich(v : ActivityNode) = 
			v.asInstanceOf[RichActivityNode]
	implicit def rich(v : ActivityNodeAspect) = 
			v.asInstanceOf[RichActivityNode]
	implicit def rich(v : ActivityEdge) = 
			v.asInstanceOf[RichActivityEdge]
	implicit def rich(v : ActivityEdgeAspect) = 
			v.asInstanceOf[RichActivityEdge]
	implicit def rich(v : Main) = 
			v.asInstanceOf[RichMain]
	implicit def rich(v : MainAspect) = 
			v.asInstanceOf[RichMain]
}
\end{lstlisting}

We can illustrate our composition mechanism by the simple Petri net example described in the previous section. 

For the class $Activity$, $Activity\-Node$ and $Activity\-Edge$,  EMF generates three Java interfaces named $Activity$, $Activity\-Node$ and $Activity\-Edge$ and three Java classes named $Activity\-Impl$, $Activity\-Node\-Impl$ and $Activity\-Edge\-Impl$. The Ker\-me\-ta compiler  does not regenerate them and simply imports them. EMF also generates a factory implementation. In addition, our Ker\-me\-ta Compiler generates Scala aspects traits $Activity\-Aspect$, $Activity\-Node\-Aspect$ and $Activity\-Edge\-Aspect$, an implicit conversion object $Implicit\-Conversion$, and new composed class  $RichActivity$, $Rich\-Activity\-Node$ and $Rich\-Activity\-Edge$ and a new factory $Rich\-Factory$ that extends generated EMF Factory to create objects extended with mixins. Note that $k2.core.Object$ is the root class of the Kermeta language, it contains methods for reflection, and all classes inherit from this class.

Listing~\ref{lst:scalaopenclass} shows an excerpt of code snippet illustrating the generated elements.

\subsection{Evaluation of the Mashup Compiler}

We evaluate the mashup compiler regarding the challenges presented in the Subsection~\ref{sec:challenges}. For the first challenge, Subsection~\ref{sec:compilation} showed how we can express the mashup composition semantics using the Scala composition operator and an architecture style that uses the Scala implicit conversion operator and the abstract factory pattern. This section focuses mainly on the composition process automation for this challenge. Then, we provide a quantitative evaluation for challenges 2 and 3. We show how the Kermeta workbench supports the integration with EMF legacy code. We discuss the performance of the resulting Scala code comparing to the Java version of the fUML case study.

\subsubsection{Is the mashup of DSLs fully automated?}
\label{claim:mashup-fully-automated}

Kermeta fully automates the mashup of the different concerns of a DSL implementation.
The language designer does not need to manipulate other constructs than \texttt{require} and \texttt{aspect} (see section~\ref{sec:mahsup}). Then, executing the mashup simply means writing:\\
\texttt{dslprogram.execute(args)}\\
(where \texttt{dslprogram} is an instance of the metamodel). 

Furthermore, Kermeta supports compiling the mashup to a standalone Java jar file.
In this case, the DSL runtime requires only a standard Java Virtual Machine and can be executed on the command line. Moreover, as a jar file, it can also be used as a library for IDE plugins (such as Eclipse plugins). This feature can be tested using the last version of the Kermeta workbench \footnote{see \url{https://gforge.inria.fr/forum/forum.php?forum\_id=11075}}.

\subsubsection{Is the mashup compatible with all components of EMF ecosystem?}
\label{claim:mashup-interoperable}

Our approach to implement DSLs as mashups of meta-languages is based on EMF/ECore metamodels. Thus, the mashup can be seamlessly integrated with all EMF based technologies. Kermeta-based DSLs can readlily benefit of all the generic and generative technologies which exist in the EMF ecosystem. In particular, it is compatible with textual or graphical syntax mappers and editor generators.

The tooling for manipulating the textual syntax of DSLs (tokenizers, parsers, editors) can be generated with textual syntax mappers. The generated tools translate some text (DSL programs) to instances of the abstract syntax. Hence textual syntax mappers are logically and technically independent of the static and the behavioral semantics and there is no need from DSL designers to provide anticipated hooks or pieces of semantics for this concern.
Note that there are already very good textual syntax mappers to provide parser and editors from text to instances of an Ecore metamodel (in the context of EMF/Ecore, for instance, EmfText\footnote{see \url{http://www.emftext.org/index.php/EMFText}} and Xtext\footnote{see \url{http://www.eclipse.org/Xtext/}}).

Along the same line, Kermeta-based DSLs can benefit from the graphical editor generator technologies existing in EMF. Again, DSL designers do not have to anticipate the possible use of a graphical syntax when they design the DSL, and this hampers by no means the possibility of later generating a graphical editor for the DSL under consideration. For instance, in the context of fUML, an editor based on UML activity diagrams can be generated to specify and modify fUML models.

\subsubsection{Quantitative Evaluation of the Execution Time}
\label{subsec:runtime}

In the previous section, we have shown that the design and the implementation of a DSL can be
elegant in terms of separation of concerns and conciseness.
However, this may come to the price of a high performance overhead.
In this section, we provide quantitative insights into the execution performance of domain-specific languages implemented using Kermeta.

More specifically, we compare a compiled version of the Kermeta-based fUML implementation with the Java-based reference fUML implementation. The compiled version was created using the Kermeta compiler. The compiler generates JVM-ready classes out of the mashup of the abstract syntax, the static semantics and the behavioral semantics.

\paragraph{Experimental Protocol}

\textbf{Subjects}
\label{subsubsec:examples}

In the following, we list the fUML programs that we use to measure the performance. We also report on their size in terms of number of model elements. The three first activities were obtained from the reference fUML implementation while the last example was created by the \emph{Logo2fUML} transformation using as input the \emph{Hilbert} model defined in the \emph{Kermeta's KmLogo Language Example}. As output, this transformation returns a well-formed fUML model of the corresponding activity.

\noindent\textbf{TestSimpleActivities:}
is one \emph{fUML} test model contained in the reference implementation. This model uses different control nodes of fUML  (\emph{ForkNode}, \emph{MergeNode}, \emph{DecisionNode} and \emph{JoinNode}). Different external activities are called by the \emph{TestSimpleActivities} executable model using a sequence of \emph{CallBehaviorAction}. 

\noindent\textbf{TestIntegerFunctions:} 
contains other activities that the previous ones and demonstrates the utilization of some primitive functions contained in the fUML library (it uses a set of \emph{CallBehaviorAction} to call these primitives functions). Size of the model: 158 model elements.

\noindent\textbf{Contacts:} 
is another example included in the reference implementation. This example is based on a simple operation that counts how many objects are inside a list and returns this number for the user. 
 The activity starts creating an object with a \emph{CreateObjectAction}. This object contains a list in which contacts can be inserted. Then, it counts how many objects are in the database. Afterwards, an object is created using the \emph{CreateObjectAction} (\emph{"create contact1"}) and added to the list using the \emph{CreateLinkAction}. Finally, the model counts again how many objects are in the database and verifies that there is one object in the database.
Size of the model: 121 model elements.

\noindent\textbf{Hilbert:}
is a program example based on the \emph{Hilbert curve} problem~\cite{Hilbert}. 
The main algorithm computes a continuous fractal space-filling curve, using a limited series of activity recursions.
Size of the model: 723 model elements.

\paragraph{Measurement}

The main metric of the experiment is the execution speed. The measurement is subject to the following characteristics:

\begin{packed_itemize}
  \item we measure only the execution time and not the loading time of models. The rationale is that the reference fUML implementation and our implementation use different loading libraries (the reference implementation uses a specialized loading library while our implementation uses the EMF loading facilities).
  \item we present the average execution time based on 30 different runs. This allows to mitigate the measurement variations due to our multitask operating system.
  \item The measurement are done on a computer with an \emph{Intel Core 2 duo 3GHz} processor and 4Gb memory, running Windows XP.
\end{packed_itemize}

We also discuss the compilation time that needs to stay reasonable to be useful in practice.

\begin{figure} [htb]
\hspace{-1em}
\includegraphics[width=\columnwidth]{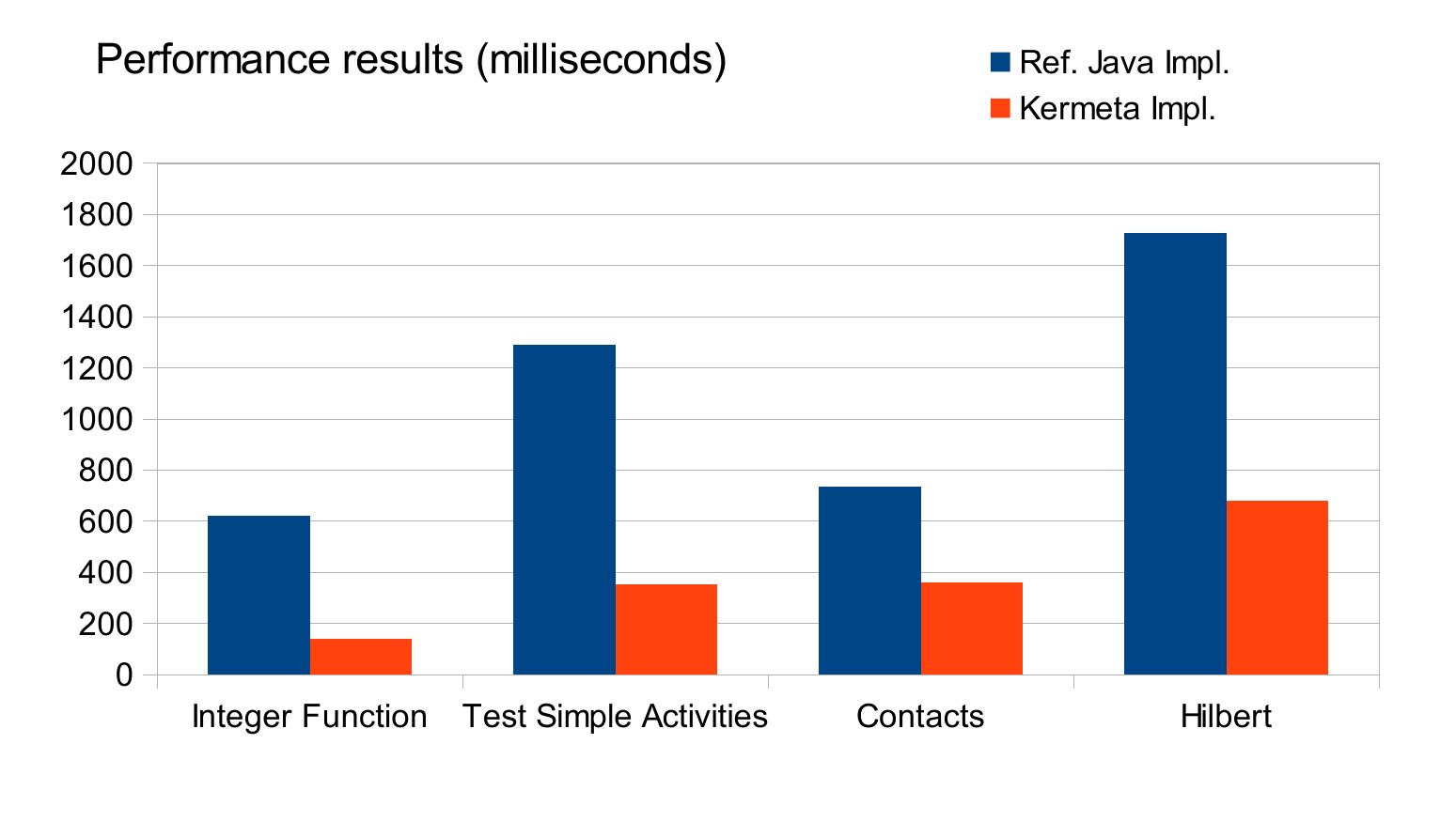}
\caption{Execution time using the  reference fUML implementation and the Kermeta-based implementation.}
\label{fig:performance}
\end{figure}

\paragraph{Results}
\label{subsubsec:runtimeResults}
Let us now present the execution time of the Java-based fUML implementation and the compiled version of our Kermeta-based fUML implementation.
Figure \ref{fig:performance} describes this comparison, presenting the mean in milliseconds of the execution time of each model. To obtain the following result, we run the Java-based fUML interpreter and the Kermeta-based one for each test model. We instrument the code to get the time before the loading of model, after the load of model and after the end of the execution of the fUML program. For sake of artifact evaluation and replication, fUML models and the Kermeta compiled version of the overall fUML metamodel are available online\footnote{\url{http://goo.gl/4RR9k}}.

A first finding is that our approach does not introduce a run-time performance overhead. Interestingly, all executable models run faster using the Kermeta-based implementation.  The ratio is not always the same. It can be explained as they use different types of actions that not all equally optimized in the Kermeta version and in the Java Version.
Note that the metrics do not take into account the loading time of models. However, we also measured the loading time of models and our Kermeta-based implementation is also faster than the one of the reference implementation. In other words, the Kermeta implementation is faster with respect to both loading and execution time. 

To understand this performance improvement, we can note that several factors have an influence on this performance improvement.  Among this factors, three of them can justify that the Java implementation is slower than the Scala/Kermeta one:
\begin{enumerate}
\item The current Java reference implementation has not been designed for performance. Some piece of the code (\emph{e.g.} to break some loops) have been designed more efficiently in the Kermeta implementation. 
\item It has been shown that Scala is more efficient than Java.  In~ \cite{citeulike:10319713}, Robert Hundt shows why Scala can run faster than Java in some cases. In particular he explains that compared to Scala,  Java also creates a temporary object for \emph{foreach} loops on loop entry. He found that this was one of the main causes of the high GC overhead for the Java version and changed the Java \emph{forall} loops to explicit while loops. This over-use of the Garbage Collector during the loop is also a main reason why Scala goes faster during the loop that the standard Java implementation.
\item Unlike the Java compiler, the Scala compiler already performs several optimisations on its own: method inlining, escape analysis (for closure elimination), and tail call optimisation. Since we compile using the \emph{-optimize} option, all these optimisations are obtained in the current bytecode obtained from the Kermeta code.
\end{enumerate}

To sum up, this experiment shows that although our approach uses and composes different meta-languages to implement a DSL, the advantages of having a clean design and a concise implementation are not traded for run-time performance overhead. These benefits to both the design-time and run-time are possible thanks to the definition of an advanced compiler considering high-level concepts as input and producing efficient code as output. Our approach also largely benefits from the large body of work on the efficiency of Scala. 

A reasonable compilation time allows easy prototyping and efficient debugging. Thus, we have measured the execution time of the Kermeta-to-Scala compilation step, and the Scala-to-bytecode compilation step on different use cases.
The compilation steps for fUML (375 classes) take 1.5 second to load the Kermeta model (the AST of the mashup),  1.8 seconds to produce the Scala code and 61 seconds to produce the bytecode. This compilation time can be a problem for designers. Indeed it can become painful if designers have to wait several seconds after each modification in the semantics to run a simulation. In investigating in which steps the compiler spends time, we can see that most of the time is consumed by the Scala compiler to type check the fUML output Scala code (20 sec),  and to optimize the bytecode (inlined in 20 sec). Since Scala is a popular language, it now provides an incremental compiler. We could improve our tool chain to use the compiler incremental feature provided by SBT\footnote{see \url{http://code.google.com/p/simple-build-tool/}}, Scala maven plugin\footnote{see \url{http://scala-tools.org/mvnsites/maven-scala-plugin/}} or FSC\footnote{see \url{http://www.scala-lang.org/docu/files/tools/fsc.html}}.

\section{Related Work}
\label{sec:rw}
 
Much work has been done on modeling languages definition, included the definition of the abstract syntax, and the static and behavioral semantics~\cite{jsw09}. Nevertheless none of them provides the possibility to combine different meta-languages as a core language design principle. For this purpose, Kermeta provides a composition operator (\emph{i.e.,} the \texttt{require} keyword) that allows us to compose the various concerns of a language, each described with a dedicated formalism. Also, the sake of composing meta-languages contains specific issues that are not well handled by generic software composition engines \cite{Harrison2006,Apel2009} (\emph{e.g.,} the composition of the contracts of the static semantics).

The mechanism provided by Kermeta with the \texttt{aspect} keyword can be compared to the open class mechanism~\cite{Clifton00multijava:modular}.
Unlike the design pattern \emph{Visitor}~\cite{Gamma:1995:DPE}, open classes do not require advance planning and preserve the ability to add new properties and methods modularly and safely. 
Open Class mechanism is useful for Kermeta, because it permits to reuse and enrich classes declared in metamodels. Indeed, in most cases, metamodels are built with domain analysis and specified using a graphical modeling editor. Static semantics or behavioral semantics is later integrated through aspects, which are statically introduced at relevant places. The composition is done statically and the composed metamodel is typed-\-che\-cked to ensure the safe integration of all units. Aspectized classes in Kermeta organize cross-cutting concerns separately from the metamodel.

Note that existing commercial frameworks perform composition of different languages through compilation. For example, all languages in the .NET framework are interoperable given their compilation to the Common Intermediate Language (CLI). In .NET, certain intra-language composition capabilities are similar to the Kermeta workbench (C\# offers a kind of open-class).  However, there is no inter-language class composition as the Kermeta workbench supports.

The modular compilation scheme we propose uses Scala as a target language. Other languages could be used to compile the features provided by Kermeta. For example, Groovy offers a reflexivity layer to dynamically extend a meta-class but clearly raises performance issues~\cite{DinkelakerEM10}. AspectJ could be also used to extend the generated classes in the compilation scheme, but Scala lets the compiler minimizes the representational gap thanks to features such as mixins, implicit, and function type.

Several authors explored the problem of modular language design (\emph{e.g.,}~\cite{Wyk2002,henriques05,vanwyk10scp,Ekman2007}). For example, LISA \cite{henriques05} and Silver \cite{vanwyk10scp} integrate specific formal grammar-based language specifications supporting the automatic generation of various language-based tools (\emph{e.g.,} compiler or analysis tools). One practical obstacle to their adoption is a perceived difficulty to adopt specific grammar-based language specifications, and the relative gap with the development environments daily used by software engineers. JastAdd~\cite{Ekman2007} combines traditional use of higher order attribute grammars with object-orientation and simple aspect-orientation (static introduction) to get better modularity mechanism. With a similar support for object-orientation and static introduction, Kermeta can be seen as a symmetric of JastAdd in the DSML world. 
Rebernak et al~\cite{Rebernak2009} and Krahn et al.~\cite{Krahn2008} 
contributed to the field in the context of model-driven DSLs. While there approaches was designed with the goal of easy composition of DSLs, they also advocate modularity of DSL compilers and interpreters. In the Kermeta workbench we left for now DSL composition as future work, but we go further on DSL modularity: we marry modularity and low representational gap, the latter being obtained thanks to appropriate meta-languages (EMOF, OCL, and the Kermeta action language).

A language workbench is a software package for designing software languages \cite{fowler2005language,volter2011programming}.
For instance, it may encompass parser generators, specialized editors, meta-languages for expressing the semantics and others \cite{Ghosh:2010,fowler:2010,mernick:2013,books/daglib/0030751}. Early language workbenches include Centaur \cite{borras1989centaur}, ASF+SDF \cite{klint1993meta}, TXL \cite{Cordy1988} and Generic Model Environment (GME) \cite{SztipanovitsK97}. 
Among more recent proposals, we can cite Metacase's MetaEdit+~\cite{tolvanen2003metaedit}, Microsoft's DSL Tools \cite{cook2007domain}, Clark et al.'s Xactium \cite{Clarka08}, Krahn et al's Monticore \cite{Krahn2008}, Kats and Visser's Spoofax \cite{Kats2010}, and Jetbrain's MPS \cite{VoelterSolomatov2010}. The important difference of Kermeta is that it is 100\% compatible with all EMF-based tools (at the code level, not only at the abstract syntax level provided by Ecore), hence designing a DSL with Kermeta easily allows reusing the rich ecosystem of Eclipse/EMF. This issue was previously addressed in the context of the Smalltalk ecosystem~\cite{RenggliGN10}.  Kermeta brings in a much more lightweight approach using one dedicated meta-language per language design concern, and providing the user with advanced composition mechanisms to combine the concerns in a fully automated way.

\section{Conclusion and Future Work}
\label{sec:conclusion}

This paper presented an approach to design and implement DSLs that is based on using and composing several meta-languages.
The key ideas behind the approach are that each language implementation concern (e.g. abstract syntax, static semantics and behavioral semantics) can easily be described in an appropriate language and that the mashup of language modules can be automatically compiled.
As a result, the task of designing and implementing DSLs can be easily modularized into loosely coupled activities, with the exception of the abstract syntax on which every other component depends. 
Furthermore; we have also shown how the Kermeta language workbench produces DSL runtimes that are al least as fast as ad hoc implementations, and still fully compatible with the EMF ecosystem.

Future work will explore (1) the use of the Kermeta language workbench for generating DSL runtimes capable of  running in constrained environments (e.g. mobile and embedded devices), (2) the use of the provided DSL modularity to support the design and implementation of language families thanks to an explicit variability modeling, and (3) another level of composition: the composition of different DSLs (\emph{e.g.,} the GEMOC Initiative\footnote{cf. \url{http://gemoc.org}}), to support a coordinated use of multiple DSLs across the development of heterogeneous aspects of a system.

\bibliographystyle{unsrt}
\bibliography{sosym-mashup}

\end{document}